\documentclass[namedreferences]{SolarPhysics}

\usepackage{solaheader}	
\newcommand{\doi}{10.1007/s11207-010-9530-7}
\usepackage{spr-sola-addons} 
\usepackage{epsfig}          
\usepackage{graphicx}        
\usepackage{url}             
\usepackage[usenames]{color}
\usepackage{algorithm,algorithmic}

\renewcommand{\vec}[1]{ {\mathbf #1} }

\newcommand{\aap}{    {\it Astron. Astrophys.}}

\newcommand{\apj}{    {\it Astrophys. J.}}

\newcommand{\solphys}{{\it Solar Phys.}}

\begin{document}

\begin{article}

\begin{opening}

\title{Automated Detection of Coronal Loops using a Wavelet Transform Modulus Maxima Method}

\author{R.T.~James~\surname{McAteer}$^{1,*}$\sep
        Pierre~\surname{Kestener}$^{2}$\sep
        Alain~\surname{Arneodo}$^{3}$\sep
        Andre~\surname{Khalil}$^{4}$     
       }

\runningauthor{R.T.J. McAteer {\em et al.}}
\runningtitle{Automated Detection of Coronal Loops}

   \institute{  $^{1}$ Marie Curie Fellow, Trinity College Dublin, College Green, Dublin 2, Ireland\\
                $^{*}$ Corresponding Author, james.mcateer@tcd.ie\\
                $^{2}$ CEA, Centre de Saclay, DSM/IRFU/SEDI, 91191 Gif-sur-Yvette, France\\
                $^{3}$ Laboratoire de Physique, \'Ecole Normale Sup\'erieure de Lyon, 46 all\'ee d'Italie, 69364 Lyon c\'edex 07, France  \\           
                $^{4}$ Dept. of Mathematics, University of Maine, Orono, ME 04469 USA\\
}

\begin{abstract}

We propose and test a wavelet transform modulus maxima method for the automated detection and extraction of coronal loops in extreme ultraviolet images of the solar corona. This method decomposes an image into a number of size scales and tracks enhanced power along each ridge corresponding to a coronal loop at each scale. We compare the results across scales and suggest the optimum set of parameters to maximise completeness while minimising detection of noise. For a test coronal image, we compare the global statistics ({\it e.g.,} number of loops at each length) to previous automated coronal-loop detection algorithms. 

\end{abstract}


\end{opening}

\section{Introduction}
     \label{intro} 

Historically, images of the Sun have always presented a myriad of features far
in advance of our physical understanding of their existence and evolution. New
data with increased sensitivity, spatial- and temporal-resolution are
continually challenging our theoretical models of the solar atmosphere. This
general statement is especially true in the case of coronal loops as observed
by the fleet of extreme ultraviolet (EUV) imagers launched since the mid
1990s. Beginning with the {\it Extreme ultraviolet Imaging Telescope} (EIT:
\opencite{del98}), this list includes the {\it Transition Region and Coronal
Explorer} (TRACE: \opencite{han99}) and recently the {\it Extreme Ultraviolet Imager}
(EUVI; \opencite{how08}) onboard the twin {\it Solar Terrestrial Earth Relations
Observatory} (STEREO: \opencite{kai08}) spacecraft. From these data, our
current models suggest these coronal loops trace out hot coronal plasma
($\approx$1MK) up to heights of around 50\,Mm above the surface of the Sun, but questions remain over their temperature and density profile, temporal evolution, and 3D structure.


Studies of these open questions are limited by the ability to extract
the loop system of interest from the hundreds of others typically
present in a solar EUV image. Furthermore each individual loop as
viewed by any current instrument is probably a collection of multiple
strands, and there will be any number of these strands overlapping
along the instrument line of sight. With the launch of STEREO, solar
physicists can now view the corona from multiple angles, enabling the
3D reconstruction of coronal loops to try to overcome this problem. The
largest problem in this reconstruction is the identification of the same
feature from each spacecraft. In the ideal world, the same feature would
be tracked in image sequences as observed from each spacecraft and the 3D
reconstruction would become a mathematically simple problem. This would
enable the scientist to proceed with studying the physical parameter
of interest. Ideally this extraction should be automated in order to
make the process instantly repeatable. From these issues, there arises
a natural requirement for the automated detection of loops based on
a statistical approach. Additionally, and perhaps most pressing, the
expected data load from the {\it Solar Dynamics Observatory} (SDO) will be
overwhelming without automated feature recognition
(\citeauthor{mca04}, 
\citeyear{mca04}, 
\citeyear{mca05b}).

\inlinecite{asc08} describe five such feature-recognition algorithms
for coronal-loop identification and apply each algorithm to an example
image from TRACE. This technique of comparing algorithms to each other,
as well as to some ground truth, is the best method for rigorous testing
of any feature recognition algorithm.  In this paper we apply a sixth such
technique to the same TRACE image as described in Section~\ref{obs}. The
algorithm is described in detail in Section~\ref{methods}. The technique
is called the 2D Wavelet-Transform Modulus Maxima (WTMM) method and was
originally developed by Arneodo and colleagues as a multifractal analysis
formalism \cite{arn00, arn03, kha06}. It has been expanded to characterize
the anisotropic nature of complex structures \cite{kha06, sno08, kha09}
and also to perform an automated and objective segmentation of image
features of interest from a noisy background \cite{kha07, cad07}. In solar
physics, the technique has been used recently to study the complexity of
solar active region magnetic fields
(\opencite{mca05a}; \opencite{con08}, \citeyear{con09}; \opencite{kes09}; \opencite{mca09}),
X-ray solar-flare emission \cite{mca07}, and in tracking coronal
mass ejections \cite{byr09}. In Section~\ref{res} we explain the natural
advantages of applying this to coronal-loop identification and compare the
global statistics against those identified in \inlinecite{asc08}. Finally,
in Section~\ref{conc} we discuss the benefits and drawbacks of the WTMM
algorithm and suggest future extensions.

\section{Observations}
     \label{obs} 

In order to directly compare our results to those previously published, we have tested our algorithm on the same sub-image as in \inlinecite{asc08}. This image was observed on 19 May 1998, 22:21:43 UT, with an exposure time of 23.172 seconds with a passband centered on  171\AA . The data were de-biased, flat-fielded, and de-spiked to remove cosmic-ray spikes from the image. The resulting image is shown with our identified loops in Figure~\ref{overlay} and presents three of the main problems with detecting coronal loops: Firstly, they are curvi-linear structures: point sources, straight lines, and areas are well behaved in the [{\it x, y}] CCD plane; curves are much more difficult. Secondly, many of the loops rise almost vertically through the solar atmosphere. The rapid density drop off in the corona (with a scale height of about 50\,Mm) results in a distinct lack of loop tops along many of these features. This makes it difficult to track from footpoint to footpoint. Thirdly there is a collection of small-scale features near the bunch of footpoints, which are typically not of interest in studies of coronal loop systems. 

\section{Method}
\label{methods}

\begin{figure}    
   \centerline{\includegraphics[width=0.8\textwidth,clip=]{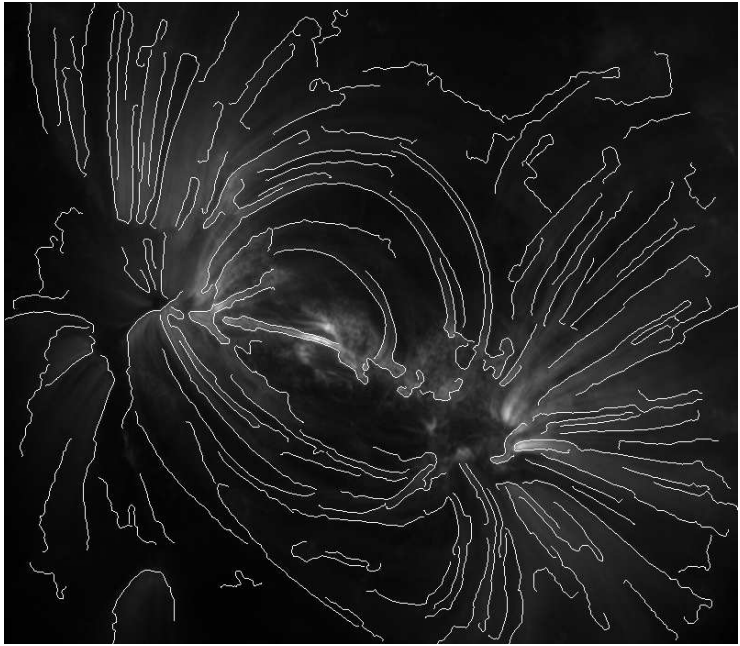}
              }
              \caption{Edge detection of coronal loops for the underlying test image. The image is $800 \times 600$ pixels, corresponding to $296 \times 222$\,Mm. Maxima chains were obtained from the wavelet transform modulus maxima, using the analyzing wavelets defined in Equation (\ref{first_order}) at a size scale of $\approx 11$ pixels.}
   \label{overlay}
\end{figure}

Our method of coronal loop identification is based on the 2D Wavelet Transform Modulus Maxima method. The continuous nature of the wavelet transform allows us to scan all size scales continuously in order to take full advantage of the space-scale information available. This allows us to perform the segmentation of objects of interest in total objectivity, without any prior knowledge on the size or morphology of the objects.

Image segmentation with continuous wavelets is based on the derivative of a 2D smoothing function (filter) acting as an ``edge detector''. Let us consider two wavelets that are, respectively, the partial derivatives with respect to $x$ and $y$ of a 2D smoothing (Gaussian) function, 
\begin{equation}
  \phi_{\rm Gau}(x,y)= \mathrm{e}^{-(x^2+y^2)/2} = \mathrm{e}^{-|{\bf x}|^2/2},
  \label{gaussian}
\end{equation}
namely
\begin{equation}
  \matrix{\psi_1(x,y) = \partial \phi_{\rm Gau}(x,y)/\partial x \hskip.1in \cr
    {\rm and} \cr \psi_2(x,y) = \partial \phi_{\rm Gau}(x,y)/\partial y.}
  \label{first_order}
\end{equation}

For any function $f(x,y) \in L^2({\bf R})$ (where $L^2({\bf R})$ consists of all square-integrable functions), the continuous wavelet transform of $f$ with respect to $\psi_1$ and $\psi_2$ is expressed as a vector \cite{mal92a, mal92b}:
\begin{equation}
  \matrix{{\bf T}_{{\bf \psi}}[f]({\bf b},a) = \cr \Bigg( \matrix{
  T_{\psi_1}[f] = a^{-2} \int \mathrm{d}^2 {\bf x}  \; \psi_1 \big(a^{-1}({\bf x} -
  {\bf b})\big) f({\bf x}) \cr T_{\psi_2}[f] = a^{-2} \int \mathrm{d}^2 {\bf x} \;
  \psi_2 \big(a^{-1}({\bf x} - {\bf b})\big) f({\bf x}) } \Bigg) \cr = \nabla \{ T_{\phi_{\rm Gau}}[f]({\bf b},a)\}  = \nabla \{ \phi_{{\rm Gau}, {\bf b}, a} * f \}.}
\label{tpsi}
\end{equation}

Thus, Equation (\ref{tpsi}) amounts to defining the 2D wavelet transform as the gradient vector of $f({\bf x})$ smoothed by dilated versions $\phi_{\rm Gau}(a^{-1}{\bf x})$ of the Gaussian filter. The wavelet transform can be written in terms of its modulus, ${\cal M}_{\bf \psi}[f]({\bf b}, a)$ and argument, ${\cal A}_{\bf \psi}[f]({\bf b}, a)$
\begin{equation}
  {\cal M}_{\bf \psi}[f]({\bf b}, a) = \sqrt{\big(T_{\psi1}[f]({\bf b},a)\big)^2 + \big(T_{\psi2}[f]({\bf b},a)\big)^2 },
  \label{modulus}
\end{equation}

\begin{equation}
  {\cal A}_{\bf \psi}[f]({\bf b}, a) = {\bf Arg} \big(T_{\psi1}[f]({\bf b},a) + \mathrm{i}T_{\psi2}[f]({\bf b},a)\big).
  \label{argument}
\end{equation}

The modulus maxima of the wavelet-transform, or intensity gradient maxima, are defined by the positions where the modulus of the wavelet transform, ${\cal M}_{\bf  \psi}[f]({\bf b}, a)$, {\it i.e.,} the gradient, is locally maximal. These WTMM are automatically organized as maxima chains which act as contour lines of the smoothed image at the considered scales. This is a non-trivial process, which has been discussed and solved empirically by \inlinecite{arn00}. At a given scale, the algorithm scans all the boundary lines that correspond to the highest values of the gradient, {\it i.e.,} the maxima chains. For each size scale considered, the algorithm outputs the [{\it x, y}]  pixel location chain of each edge detected. Each chain then corresponds to a single extracted edge in the image. We identify each edge as a single coronal loop, as shown in Figure~\ref{overlay}. This process is repeated continuously at all scales. Futher detail on the actual code is given in the Appendix.

\section{Results}
\label{res}

In this section we describe the nature of the WTMM method to trace out and connect edges as one of the major advantages of this method. We compare automatically detected loops against those manually identified in \inlinecite{asc08}. In order to quantitatively compare this technique against those previously published we also collate a number of global image statistics: the total number of loops detected is an indication of the completeness of the algorithm: the maximum loop length is an indication of the upper limit on the detection: the scaling index of the cumulative distribution shows how the algorithm performs across size scales. 

\subsection{Loop Tracing}

\begin{figure}    
   \centerline{\includegraphics[width=0.8\textwidth,clip=]{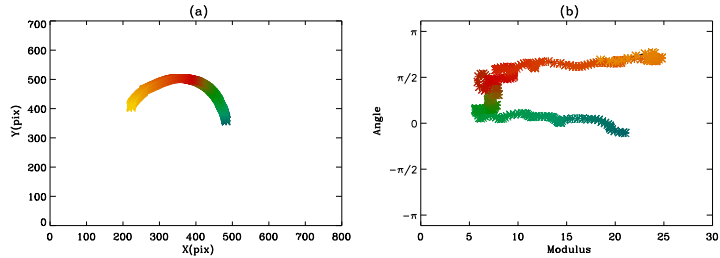}
              }
              \caption{A complete coronal loop. (a) The[{\it x, y}]  location of each pixel along the loop from one footpoint. The colour scheme traces the loop from the first footpoint (green), through the loop top (red) to the second footpoint (yellow). The axis are in pixel units, where one~pixel $\approx$ 370~km. (b) The Modulus [Equation~(\ref{modulus})] and Angle [Equation~(\ref{argument})] information at each point along the loop with the same colour scheme as in (a), where a positive (negative) angle corresponds to a counter-clockwise (clockwise) rotation from $0$ (along the positive {\it x}-direction) }
   \label{fig4}
\end{figure}

\begin{figure}    
   \centerline{\includegraphics[width=0.8\textwidth,clip=]{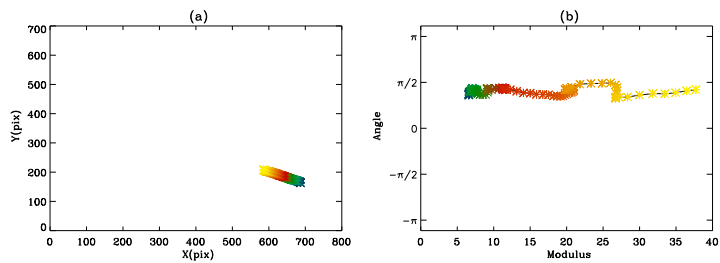}
              }
              \caption{An incomplete coronal loop. (a) The [{\it x, y}]  location of each pixel along the loop from one footpoint. The colour scheme traces the loop from the first footpoint (yellow), up through the atmosphere (red, green) where it presumably connects to another footpoint, possibly elsewhere in the image. The axis are in pixel units, where one~pixel $\approx$ 370~km. (b) The Modulus [Equation~(\ref{modulus})] and Angle [Equation~(\ref{argument})] information at each point along the loop with the same colour scheme as in (a), where a positive (negative) angle corresponds to a counter-clockwise (clockwise) rotation from $0$ (along the positive  {\it x}-direction) }
   \label{fig5}
\end{figure}

Figure~\ref{fig4}a shows an individual loop corresponding to one of the large complete coronal loops near the center of the image. The modulus and angle information used to detect this loop is shown in Figure~\ref{fig4}b with the same color scale as Figure~\ref{fig4}a. As the loop is traced from one footpoint to the other, both the modulus and angle information change smoothly. The modulus is strongest at the footpoints (where the intensity in the original image is strongest) and drops off by a factor of five at the loop tops. The angle starts off at $0\,^{\circ}$ at the right--most footpoint ({\it i.e.,} pointing to the right of the image, along the {\it x }-axis), increases to $\pi/2$ at the loop top ({\it i.e.,} pointing straight up along the {\it y }-axis), and increases to $\pi$ at the second footprint ({\it i.e.,} pointing to the left, along the {\it x }-axis). It is not a completely smooth transition; this is most evident at the loop top, where the angle varies more rapidly. This variation of angle and modulus hence provides two parameters for loop identification. Most importantly, although the modulus may drop off dramatically at the loop top (a manifestation of the decreased signal), the variation in angle is much smoother. A second example, for tracing a partial loop, is displayed in Figure~\ref{fig5} for a loop leg in the lower right of the image. In this case, the angle stays constant ($\approx\pi/2$) along the leg, while the modulus drops off with increasing distance from the footpoint.

\subsection{Comparison of Loop Locations}

\begin{figure}    
   \centerline{\includegraphics[width=0.8\textwidth,clip=]{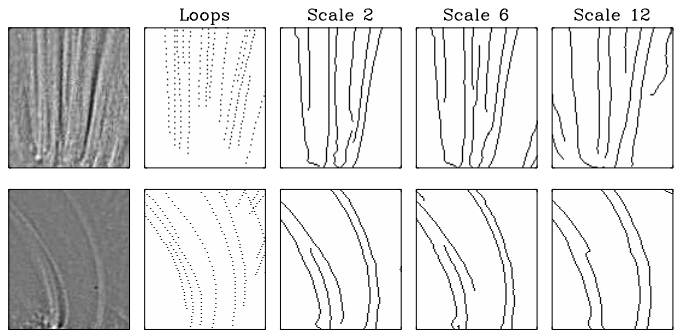}
              }
              \caption{Comparison of (left to right) the smooth-subtracted data, manually identified loops, edges from scale 2 of the WTMM, edges from scale 6 of WTMM, edges from scale 12 of WTMM. Top row is for a subimage extracted at the leftmost footpoints of the loop system consisting of mostly vertical loops, bottom row for a subimage extracted near the rightmost footpoints of the system and consists of more curved loops. Each sub-image is $100 \times 150$ pixels ($37\, 000 \times 55\, 500$~km)}
   \label{loop3}
\end{figure}

A more detailed analysis of extracted loops can be carried out by comparing the edges against those loop manually identified in \inlinecite{asc08}. Figure~\ref{loop3} shows such a comparison for two subimages of the data. These regions are picked as they illustrate two important spatial regimes of loop identification and demonstrate some of the main positive aspects and drawbacks of the WTMM method. Figure~\ref{loop3} (top) contains a bunch of loop legs. These are mostly linear and display a relatively constant intensity along each loop inside this window. These two aspects should make these features easier to detect however none of the WTMM scales completely extract the manual coordinates. Each of the scales displayed pick out seven\,--\,eight of thirteen manually detected loops. The largest scale displayed (scale 12; right column) is unable to pick out any of the weaker, shorter, loops however does produce smooth edges. The smallest scale (scale 2; middle column) identifies more of the smaller loops, but at the expense of producing a more ragged edge. Scale 6 still contains many of small loops, but also produces a smooth edge. Scale 6 is also the only scale which captures the loops in the bottom right of this subimage. 
        Figure~\ref{loop3} (bottom) shows a selection of the curvi-linear features. As these are curves in [{\it x, y}] pixel space and show a drop in intensity, they are normally more difficult to extract. However the ability of the WTMM method to chain in angle space assists in overcoming this problem. Of the seven curved, manually-detected features, each scale reproduces four\,--\,five. The large, but weak, loop near the  center of the subimage is not detected at any scale. The wavelet transform power lies below the threshold for idenfication. Lowering our threshold to extract this loop results in pulling out an unmanageable number of weak non-loop features. Scale 2 and scale 6 are similar; the main difference is the ability of scale 6 to begin to detect the loop in the top left. At scale 12, the smoothing window is so large that the tracking algorithm jumps from one large loop to a second large loop. Clearly there is a trade off to be made between completeness and quality of extracted edges. 

\subsection{Number of Loops and Maximum Loop Length}

\begin{figure}    
   \centerline{\includegraphics[width=0.8\textwidth,clip=]{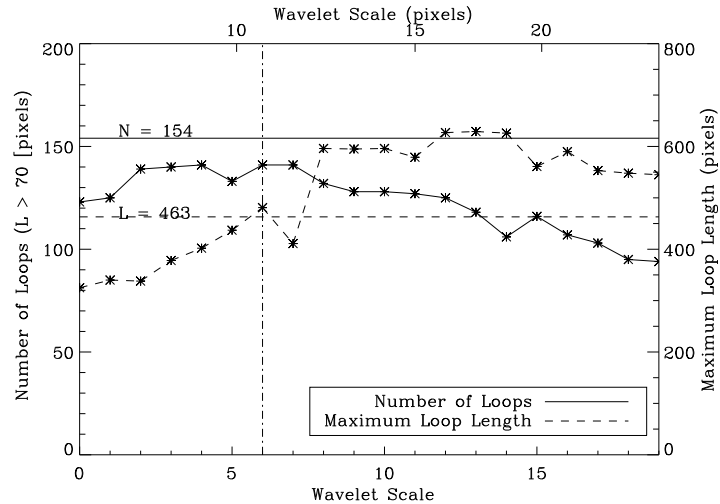}
              }
\caption{The completeness and upper limits across each wavelet scale. The
bottom axis corresponds to the wavelet scale order; the top axis is the
same value converted to pixel units. The total number of loops with size
greater than 70 pixels [$N$], is plotted as a solid line and asterisks
according to the left axis. The solid horizontal line is plotted at
the desired complete value of $N=154$. The maximum loop length
[$L$] is plotted as a dashed line with asterisks according to the right
axis. The dashed horizontal line is plotted at the desired upper limit
of $L=463$. Both desired values are from the manually identified loops
in Aschwanden {\it et al.} (2008)}.

   \label{fig2}
\end{figure}

A full quantitative comparison of the positive and negatives of each scale is best achieved by comparing the global statistics. The total number of detected loops and maximum loop length at each wavelet scale size is displayed in Figure ~\ref{fig2}. The bottom axis shows the wavelet scale, with the corresponding scale in pixels displayed on the top axis. The solid line shows how the total number of loops greater than $70$ pixels (left axis) varies with each wavelet scale size. The horizontal solid line shows the ground truth ({\it i.e.,} manually extracted)  of $N=154$. It is clear that the wavelet algorithm consistently underperforms, but it is noticeable that the plot peaks at scale size $6$. At this scale we can reach a completeness ratio of $0.91$. The dashed line shows how the maximum loop length (right axis) varies with each scale size. The horizontal dashed line shows the ground truth ({\it i.e.,} manually extracted) value of $L=463$. As expected, the ability of the algorithm to detect the largest loops increases with increased wavelet scale size. It underperforms at small scale size, and over-detects larger structures at large scales sizes. The best performance is also at wavelet scale 6, where we reach a detection ratio of $1.05$. It is comforting to note that we achieve maximum accuracy and completeness in these two parameters at the same scale size. 

\subsection{Scaling Index of Cumulative Frequency Distribution}

\begin{figure}    
   \centerline{\includegraphics[width=0.8\textwidth,clip=]{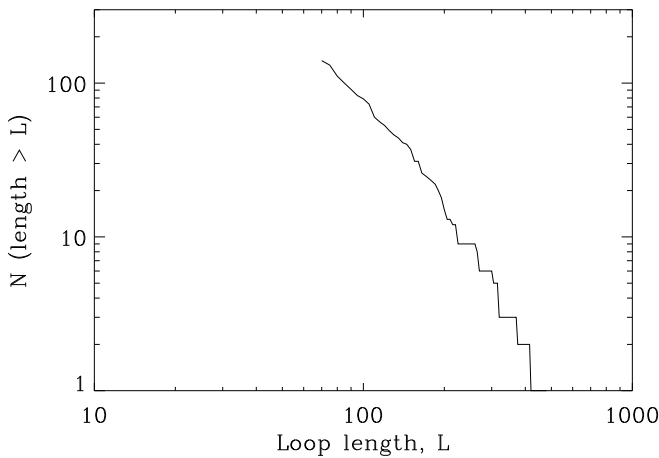}
              }
              \caption{The cumulative distribution of loops with length greater than 70 pixels}
   \label{fig3}
\end{figure}

The cumulative distribution of loops with length greater than 70 pixels is displayed in Figure~\ref{fig3} for scale 6 ($\approx 11$ pixels). The scaling index ($\beta$) of this plot is indicative of the distribution of loop lengths in the image and reflects the tendency of smaller loops of naturally outnumber larger loops. \inlinecite{asc08} report a ground truth of $\beta = -2.8$ for the manually traced loops, with the five algorithms producing values between -2.0 and -3.2. For the WTMM algorithm in Figure~\ref{fig3}, $\beta = -2.78 \pm 0.1$.

\section {Conclusions and Future Work}
\label{conc}

EUV images of the corona provide a multitude of information regarding the plasma properties of coronal loop systems. A number of algorithms currently exist that attempt to extract the locations of these loops automatically. These algorithms are discussed in detail in \inlinecite{asc08} but here we compare a brief outline of each one against our WTMM method. 

The Oriented Connectivity Method (OCM: \citeauthor{lee06a}, \citeyear{lee06a}) consists of
a preprocessing step to remove non-loop candidates, a linkage step based
on magnetic-field extrapolation guide, and post-processing to spline
fit and link loop segments. The Dynamic Aperture-Based Loop Segmentation
Method (DAM: \citeauthor{lee06b}, \citeyear{lee06b} ) replaces the linkage step of the OCM with a
search for loops by connecting pixels which have a similar Gaussian cross
sectional profile \cite{car03} and orientation. The Oriented-Directivity
Loop Tracing Method (ODM: \citeauthor{asc08}, \citeyear{asc08} ) is essentially a local-directivity
version of the OCM, tracing loops locally in two directions from their
tops to their footpoints. The Ridge Detection by Automated Scaling
(RAS: \citeauthor{inh08}, \citeyear{inh08} ) is is a multiscale ridgel extension of the
OCM. Finally, the Unbiased Detection of Curvi-Linear Structures
(UDM: \citeauthor{ste96}, \citeyear{ste96} ) is a more generic method consisting the determination of the
centroid of the loop structures from a second derivative perpendicular
to the loop, and extended \cite{rag04} to connect structrues using a
generalised Radon transform. \inlinecite{asc08} show that the OCM and
DAM both successfully extract the large loop features. The ODM, RAS,
and UDM codes contain many more free parameters and extracted more
segmented small loop-like features.

The WTMM method presented here offers many natural advantages over other techniques: Firstly, it has naturally directional linkage. The algorithm works by tracking along edges (perpendicular to the gradient) hence the resulting edges require minimum post-processing linkage. Secondly it is naturally multiscale, enabling a scientist to extract features in the size-scale range of interest. Essentially we negate the need for preprocessing by simply studying the scale at which the feature of interest occurs. These two features combine to provide a good localisation and a single response to each edge in the image. These advantages are evident in the good comparison of global statistics against existing algorithms from \inlinecite{asc08}. We reach a completeness of 0.91 and detection ratio of 1.05, with a scaling index for the cumulative distribution of $\beta = -2.78 \pm 0.1$. These three statistics compare favourably to those in \inlinecite{asc08}. The WTMM contains a few free parameters which we have attempted to optimise for loop detection. The most important free parameter is the choice of the form of the mother wavelet. We choose the Derivative of Gaussian (DOG) as it is well studied mathematically and has previously been successful in tracking edges. We note that other, more naturally curvi-linear multiscale algorithms exist ({\it e.g.,} curvelets, ridgelets) which may assist in providing better identification of individual coronal loops. The other main free parameters are the choice of thresholds in tracking the edges (see Appendix for details). Finally we propose that our sixth wavelet scale (corresponding to $\approx 11$ pixels) seems to optimise our ability to pick out the small scale features, retain smoothness in the extracted coordinates, and best agrees with the expected global statistics

An obvious extension of this work is to use the multiscale aspect in a soft-thresholding sense: currently we decide on an optiumum scale, there is the possiblity to instead decide on a best scaling index. This would consist of attempting to track each feature across scale. A noisy feature is expected to contain a lot of power at small scales, with very little power at larger scales. For a real feature, particularly the large complete coronal loops, there is expected to be much less change across scales. This may assist in removing non-loop features near the footpoints of the loops system. We also note our extracted loops, even at scale 6, are probably too ragged for stereoscopy and a degree of post-processing smoothing may still be neccessary. When applied to EUVI images from both STEREO ahead and behind data, this may allow for a 3D reconstruction of these loop systems with less manual labour. We expect algorithms such as the one studied here will be vital in the SDO era for the tracking of loops across a sequence of images, and hence in the expanding field of coronal seismology.

%
\appendix   
\section{Light Weight Notations for WTMM Formulae}

Let $f(x,y)$ be the input image to be analized and $T$ the wavelet transform vector of $f$ at scale $a$. The wavelet transform components are 
\begin{eqnarray}
f_x &= &\partial_x (f \star \phi_a)\\
f_y &= &\partial_y (f \star \phi_a)
\end{eqnarray}
The square modulus of the wavelet transform vector is $M^2=f_x^2+f_y^2$.
Let us note higher order derivative of $f$ this way using multiple $x$- or $y$-indices:
$\partial_x f_x = f_{xx}$, $\partial_y f_x = f_{xy}$, {\it etc...}

\section{WTMM Edge Definition}

The WTMM are defined as the locations of the points where there is a maximum of the wavelet transform modulus along the direction of the wavelet transform vector, {\it i.e.,} WTMM are the location of the greatest slope in the $f \star \phi_a$ landscape. The steepest slope line is not exactly orthogonal to the WTMM edge (only in particular cases). To find those points, we need to evaluate the scalar quantity ($N$) defined as the dot product 
\begin{equation}
  N = \vec{\nabla}(M^2)\bullet\vec{T}
\end{equation}
At each WTMM location, $N$ is zero and $N$ changes its sign when moving along
the direction $T$ and crossing the WTMM. This is not enough to clearly
identify a maximum, we also require that the second derivative along the direction $T$ must be strictly negative, {\it i.e.},
\begin{equation}
  N' = \partial_x N  f_x + \partial_y N f_y < 0
\end{equation}
The quantities $N$ and $N'$ are computed exactly:
\begin{equation}
  N = \vec{\nabla}(M^2)\bullet\vec{T} = 2f_x^2f_{xx} + 4f_xf_yf_{xy} + 2f_y^2f_{yy}.
\end{equation}
$N'$ is given by:
\begin{eqnarray*}
  N' = \partial_x N f_x + \partial_y N f_y &=& 4f_x^2f_{xx}^2+4f_y^2f_{yy}^2+2f_x^3f_{xxx}+2f_y^3f_{yyy}\\
& &+4f_x^2f_{xy}^2+4f_y^2f_{xy}^2\\
& &+ 8f_xf_yf_{xy}f_{yy} + 8f_xf_yf_{xy}f_{xx}\\
& &+6f_x^2f_yf_{xxy}+6f_xf_y^2f_{xyy}
\end{eqnarray*}

\section{WTMM Computation Algorithm}

Finally, the full algorithm is summed up. It is essentially a FOR loop over pixel location. Note that the final edge image is made of the pixels containing a WTMM. The modulus at the ``exact'' WTMM location is adjusted using a polynomial fit using points along direction $\vec{T}$.
 \begin{algorithm}[H]
   \caption{\textit{WTMM edges computation} algorithm}
   \begin{algorithmic}
     \REQUIRE $f(i,j)$ the input image
     \REQUIRE $a$ scale parameter
     \STATE compute WT of $f$ at scale $a$ : $\vec{T} = \vec{\nabla}(f \star G_\sigma)$
     \STATE compute $N$ and $N'$
     \FOR {pixel $(i,j) \in $ image range}
     \IF{$N'<0$ and $N$ changes sign in $3\times3$-neighbourhood}
     \STATE pixel $(i,j)$ is labelled as a WTMM
     \STATE perform a third order polynomial interpolation along direction $\vec{T}$ to get an accurate wavelet transform modulus value at maximum
     \ENDIF
     \ENDFOR
     \\WTMM edge image
   \end{algorithmic}
 \end{algorithm}
Several free parameters have been statistically and/or empirically adjusted over the years. The following have been used for this study:

\noindent - {\it Spatial resolution: Finite-size and edge effects}. The minimum scale at which the wavelet transform is still resolved is seven pixels. On the other hand, in order to avoid artificial effects from the edges of the images, only the central 72\% of the original wavelet transformed image should be kept for analysis. A methodical calculation of this parameter is carried out in  \inlinecite{arn00}

\noindent - {\it Scale resolution}. For an image of 1024 $\times$ 1024 pixels, usually 50 wavelet scales are considered, from seven pixels to $\approx$ 200 pixels, in $\log_2$ steps ({\it i.e.,} very high resolution for small scales and low resolution at high scales). The seven-pixel lower limit restricts our resolution and is adopted to reduce the computational overhead.

%
\begin{acks}

RTJMCA is a Marie Curie Fellow funded under FP6. The authors thank an anonymous referee for many useful comments. This work is made possible due to many fruitful discussions at the Solar Image Processing Workshop series. 

\end{acks}



\IfFileExists{\jobname.bbl}{} {\typeout{}
\typeout{****************************************************}
\typeout{****************************************************}
\typeout{** Please run "bibtex \jobname" to obtain} \typeout{**
the bibliography and then re-run LaTeX} \typeout{** twice to fix
the references !}
\typeout{****************************************************}
\typeout{****************************************************}
\typeout{}}

\end{article} 
\end{document}